\documentstyle[aps,preprint]{revtex}
\begin{document}
 
\draft
\title{Pion Content of the Nucleon as seen \\
      in the NA51 Drell-Yan experiment}
 
\author{H. Holtmann ${}^1$, N.N. Nikolaev ${}^{1,2}$,
J. Speth ${}^1$ and A. Szczurek ${}^3$}
\address{
${}^1$ Institut f\"ur Kernphysik, Forschungszentrum J\"ulich, \\
52425 J\"ulich, Germany \\
${}^2$ Landau Institute for Theoretical Physics, \\
GSP-1, 117940, ul. Kosygina 2, 117334 Moscow, Russia \\
${}^3$ Institute of Nuclear Physics, PL-31-342 Krak\'ow, Poland}
\date{\today}
\maketitle
 
\begin{abstract}
In a recent CERN Drell-Yan experiment the NA51 group found a strong
asymmetry of $\bar u$ and $\bar d$ densities in the proton at
$x\simeq0.18$. We interpret this result as a decisive confirmation of
the pion-induced sea in the nucleon.
\end{abstract}
 
 
With the advent of high precision data on deep inelastic scattering,
understanding of the nonperturbative flavour structure of the
nucleon is becoming one of the pressing issues where the interests
of particle and nuclear physics converge.  At large $Q^2$ the
perturbative QCD evolution is flavour independent and, to leading
order in $log\>Q^2$, generates equal number of $\bar u$ and $\bar d$
sea quarks. The $\bar u-\bar d$ asymmetry coming from effects of the
interference between the $u$ and $d$ quarks from the perturbative
sea $\bar uu$ and $\bar dd$ pairs and the valence $u$ and $d$ quarks
of the nucleon was found to be negligible \cite{RS79}. Perturbative
QCD describes only the $Q^2$-evolution of deep inelastic structure
functions starting with certain nonperturbative input. There are no
a priori reasons to expect a $\bar u$-$\bar d$ symmetric
nonperturbative sea. Furthermore, the strong correlations between
quarks and antiquarks of the nonperturbative sea, exemplified by the
pionic field in physical nucleons, leads to precisely such an
asymmetry \cite{HM90,SST91,HSB91,Z92,SS93,SH93,HSS'93}.  There is
already experimental evidence for such an asymmetry from the
Gottfried Sum Rule violation observed by the NMC collaboration at
CERN \cite{A91}.
 
In the simplest models the nucleon is treated as a system of three
quarks. The pion-nucleon interaction leads naturally to an admixture
of a $\pi N$ Fock component in the physical nucleon. In the simplest
approximation, the Fock state expansion of the light cone proton
reads (shown are the probabilities of the different components):
\begin{equation}
|p\rangle_{phys}={1\over Z}\Big[|p\rangle_{core}
         +n_\pi( {1\over 3} |p\pi^0\rangle
               + {2\over 3} |n\pi^+\rangle)\Big]
\end{equation}
with $Z$ being the wave function renormalization constant defined by
$Z=1+n_\pi$. Here $n_\pi/Z$ gives the probability to find the
physical proton in a light cone Fock state consisting of a nucleonic
core and a pion.
 
The $\pi_0$ quark distributions are symmetric functions of up and
down quarks.  The quark content of the $\pi^+=u\bar d$ implies,
however, that $\bar u<\bar d$. Recently the CERN NA51 collaboration
\cite{K94} presented the first direct measurement of the ${\bar u /
\bar d}$ asymmetry
\begin{equation}
{\bar u(x)\over \bar d(x)}
   =0.56\pm0.04\pm0.05\qquad {\rm at}\ \ x=0.18.
\label{expr}
\end{equation}
In this paper we interpret the NA51 result as decisive evidence for
the pion induced nonperturbative sea in the nucleon.
 
Before the NA51 result, the experimental evidence for any $\bar
u-\bar d$ asymmetry came from an analysis of the Gottfried-Sum-Rule
(GSR) violation \cite{A91}, which in the Quark-Parton-Model can be
related to
\begin{equation}
GSR=\int_0^1 \; \bigl[ F_2^p(x)-F_2^n(n) \bigr] \; {dx\over x}
   ={1\over3}+{2\over 3}\int_0^1 \; \bigl[\bar u(x)-\bar d(x) \bigr]
  \; dx \; .
\end{equation}
Experimentally, the determination of $F_2^n$ is biased by the
uncertainties because of the nuclear shadowing effects at small
$x<x_{shad}\simeq 0.05$ \cite{NZ92,BGNPZ94}. Furthermore, the GSR
analysis gives only the integrated asymmetry, and does not identify
the region of $x$ from which the asymmetry comes. Different
dynamical models predict a $\bar u/\bar d$ asymmetry concentrated in
different regions of $x$.  In this paper we wish to emphasize that
$x \simeq 0.2$ is the region where the nucleon sea is expected to be
dominated by the nonperturbative pion-induced sea.
 
The NA51 analysis is based on the comparison of the Drell-Yan
dilepton pair production in $pp$ and $pn$ collisions.  Here, one can
kinematically select annihilation of the valence quarks in the
projectile proton and the sea quarks of the target nucleon. Since
the valence quark distributions are fairly well known, one can
extract the $\bar u /\bar d$ asymmetry \cite{ES91,G92,SSG93}.  In
order to extract the $pn$ cross section, the NA51 experiment used a
deuteron target, but $x\simeq 0.2$ is a region free of the shadowing
effects.
 
Ever since the observed violation of the GSR, there were attempts to
accomodate the GSR-implied $\bar u/\bar d$ asymmetry into the QCD
evolution analysis of nucleonic structure functions. Usually the
asymmetry is introduced in terms of plausible parametrizations and,
depending on the parametrization, the asymmetry can be placed either
in the large-$x$ or small-$x$ region, reflecting one's prejudice
about where the violation of the GSR comes from. As an example we
will present the $\bar u /\bar d$ ratio using the
Martin-Stirling-Roberts parametrization \cite{MSR93}($D'_0$,
$D'_-$), which was fitted to the previously available data on deep
inelastic scattering and Drell-Yan dilepton production. Due to the
functional form of the parametrization employed, this fit
implicitely assumes that the $\bar u$ - $\bar d$ asymmetry is
concentrated in the small $x$ region. We also show predictions from
the model of the pion induced sea which is in a good agreement with
the NA51 result \cite{K94}. Below we describe the corresponding
calculation in more detail.
 
The pionic corrections to the quark distributions of the physical
nucleon are given by the convolution formula
\begin{eqnarray}
q_{p,phys}(x)={1\over Z}\Big[q_{p,core}(x)
&\displaystyle
  +\int_x^1 n_\pi(1-y)\Big({1\over3}\>q_{p,core}({x\over y})
            +{2\over3}\>q_{n,core}({x\over y})\Big){dy\over y} 
             \nonumber\\
&\displaystyle
  +\int_x^1 n_\pi(y)\Big({1\over3}\>q_{\pi^0}({x\over y})
            +{2\over3}\>q_{\pi^+}({x\over y})\Big){dy\over y}
                        \Big].
\label{conv}
\end{eqnarray}
The function $n_\pi(y)/Z$ (for formulas see \cite{Z92,HSS93}) gives
the probability that the physical proton is in the virtual
'nucleonic core'-meson state with the meson carrying a longitudinal
momentum fraction $y$ of the proton momentum. Thus, $n_\pi(1-y)/Z$
is the probability that the corresponding nucleonic core carries a
longitudinal momentum fraction $y$.  Remarkably enough, the number
of pions $n_\pi$ and the $\bar u$-$\bar d$ asymmetry can be
calculated essentially parameter-free using pion-nucleon interaction
parameters inferred from low- and high-energy soft hadronic
interactions \cite{Z92,SH93,HSS93,HSS'93}.  Specifically, the
pion-exchange contribution enters the high energy nucleon (delta)
production in hadronic reactions (Fig.1a) and the lepton deep
inelastic scattering off virtual pions (Fig.1b) in similar
kinematical conditions. Both processes are described by the same
light cone $\pi N$ wave function, which allows one to constrain the
principal parameters, the radii of the meson-baryon light cone wave
functions, from an analysis of the $p\rightarrow n, \Lambda$,
etc. fragmentation spectra \cite{Z92,HSS'93}.  The formulas for the
meson-baryon splitting functions and the parameters needed for the
analysis (coupling constants, radii of the light cone meson-baryon
Fock states) can be found elsewhere \cite{Z92,HSS93,HSS'93}.  This
model can easily be extented to include other, important Fock
components.  The full model \cite{SS93,HSS'93} includes all octet
and decuplet baryons, as well as the pseudoscalar and vector mesons.
 
In order to make predictions for the sea quark distributions, one
needs input for the $q$ ($\bar q$) distributions of the pion
(mesons). Fortunately enough, in the region of interest, the meson
induced sea is dominated by the valence quarks of the pion.  Indeed,
typically $n_{\pi}(y)$ has a maximum at $y \simeq 0.3$ and the $x =
0.18$ studied in the NA51 experiment corresponds to the rather large
Bjorken variable in deep inelastic scattering off pions $x_{\pi}
\simeq x/y \simeq 0.6$.  Thus, the recent parametrization
\cite{SMRS92} based on the $\pi N$ Drell-Yan data can be regarded as
reliable and the meson-induced sea can be calculated with good
accuracy. We assume the same parametrization to hold also for
vector mesons.
 
The primordial sea distribution of the nucleonic core is poorly
known.  One cannot use here the conventional parametrizations which
implicitly contain the mesonic effects. We adopt the following
phenomenological procedure. We assume the primordial sea in the
nucleonic core to be $\bar u$-$\bar d$ symmetric. For the shape of
the primordial sea we take the symmetric MRS ($S'_0$) \cite{MSR93}
parametrization.  Then the MSR distribution is multiplied by a
rescaling factor to obtain rough agreement with the absolute
normalization of the MRS($D'_0$, $D'_-$) \cite{MSR93} $\bar u$ and
$\bar d$ distributions. Such a procedure suggests that the
primordial sea is about 50\% of that given by the symmetric MRS
($S'_0$) parametrization.  The procedure is somewhat crude since the
($D'_0$, $D'_-$) parametrization does not reproduce the NA51 result,
and this comparison suggests a primordial sea which is somewhat
steeper than given by the MSR ($S'_0$) parametrization. But as we
shall see, in the region of interest the effect of the primordial
sea is actually a correction.
 
The result of our calculation for the $\bar u/\bar d$ ratio is shown
in Fig.2.  The solid line shows our prediction using the full model.
The decomposition of the $\bar u(x)$ and $\bar d(x)$ distribution
into contributions from different mechanisms is shown in Fig.3.  As
seen from the figure, the contribution of the $\pi N$ component to
$x\bar d$ is significantly larger than to $x\bar u$. The source of
this difference is the valence quark distribution of the
$\pi^+$. This is our main source of the $\bar u$-$\bar d$ asymmetry.
Also the $\rho N$ component (not shown separately in Fig.3)
introduces the asymmetry. The other components ($\pi\Delta$ and
$\rho\Delta$) dilute the asymmetry; because of the isospin $3/2$ of
the $\Delta$, they generate more $\bar u$ than $\bar d$ quarks.
Small contributions from the hyperon--strange meson Fock states are
also included. The sensitivity of our results to the size of the
primordial sea can be seen from Fig.2 (dashed lines), where we also
show the $\bar u(x)/\bar d(x)$ ratio with drastically different
reduction factors: 0.25 - dashed curve and 0.75 - dot dashed
curve. As can be seen from the figure, the resulting asymmetry is
quite insensitive to the reduction factor.  For comparison we also
show $\bar u/\bar d$ ratio obtained from the MSR ($D'_0$, $D'_-$)
parametrizations \cite{MSR93}. In the present approach, an absolute
normalization of the meson-induced sea is based on the combined
description of deep inelastic and hadronic fragmentation processes
of Fig.1 using the same light cone wave function. The absorption
corrections in both processes may be different, but their overall
effect is known to be numerically small, $< 10$-$20\%$
\cite{ZS84}. They suppress the hadronic reaction cross section,
which may lead to a slight underestimation of the meson-induced sea
in our calculation.
 
The quark distributions in Figs.2 and 3 have been calculated at the
momentum scale $Q^2=4\>GeV^2$. The QCD evolution to the region of
$Q^2=10$-$20\> GeV^2$, appropriate for the NA51 Drell-Yan data, has
a negligible effect on the $\bar u / \bar d$ ratio shown in Fig.2.
 
Summarizing, the new NA51 result $\bar u / \bar d = 0.56 \pm 0.04
\pm 0.05$ at $x = 0.18$ can naturally be explained by the presence
of $\pi(\rho)-baryonic\ core$ Fock components in the nucleon wave
function.  The observed asymmetry can, in principle, be reproduced
by suitably modified parametrizations which allow for stronger
asymmetry placed at somewhat larger $x$ compared to the MSR
$D'_{0}$, $D'_{-}$ parametrizations \cite{MSR93}. We emphasize that,
in contrast to parametrizations, our results are predictions from a
dynamical model which predicts nonperturbative sea distributions
without free parameters. Furthermore, our model of the nucleon makes
a link between low energy meson-nucleon couplings, high energy soft
hadron-hadron collisions and hard deep inelastic scattering of
leptons.  It is worth mentioning here that a recent lattice QCD
calculations gave evidence for the importance of pion loop effects
for nucleon properties \cite{CL93}.  New experiments at larger $x$
are neccessary to give more insight into the pionic (or more
general: mesonic) cloud of the nucleon.  The definite prediction of
the meson-induced sea model is that the $\pi N$ and $\rho N$ Fock
components dominate the antiquark distributions in the large $x$
region. Consequently, the $\bar u / \bar d$ ratio is predicted to
decrease towards $1/5$ at $x\rightarrow 1$. Unfortunately the
predictive power of the model deteriorates close to the kinematical
limits of $y \rightarrow 1$ and $y \rightarrow 0$, where the
meson-baryon light cone wave function is not well known.  The new
experiment planned at Fermilab \cite{G92} will be useful in this
respect and should provide a possibility of pinning down the $x$
dependence of the $\bar u$ - $\bar d$ asymmetry up to $x = 0.4$.
 
\vskip 1cm
 
$\it {Acknowledgements.}$ This work was supported in part by the
Polish grant 2 2409 9102.  We are indebted to B. Gibson for reading
the manuscript and interesting discussion.

\begin{figure}
\caption{
The diagrams for high energy nucleon(delta) production in a
hadron-nucleon collision (a) and the deep inelastic scattering off
virtual pion (b).}
\end{figure}
 
\begin{figure}
\caption{
The $\bar u (x) / \bar d (x)$ ratio measured by the Drell-Yan NA51
group compared to: the experimental $D'_0$ (dotted) and $D'_-$
(double-dotted) MSR parametrizations
\protect\cite{MSR93}, and our full model (solid). In addition
we show the result with the primordial sea contribution in baryons
reduced by a factor 0.25 (dashed) and 0.75 (dot-dashed).}
\end{figure}
 
\begin{figure}
\caption{
The contributions of various Fock states to the $x\bar u$ (Fig. 3a)
and $x\bar d$ (Fig. 3b) distributions.  The dotted curves correspond
to the asymmetric quark distributions $D'_0$ (dotted), $D'_-$
(double-dotted) from Ref.\protect\cite{MSR93}.}
\end{figure}
 
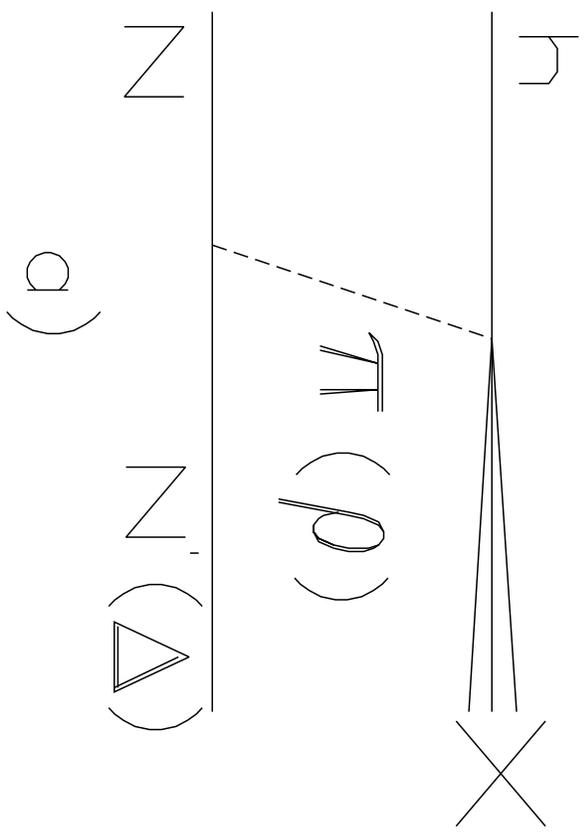
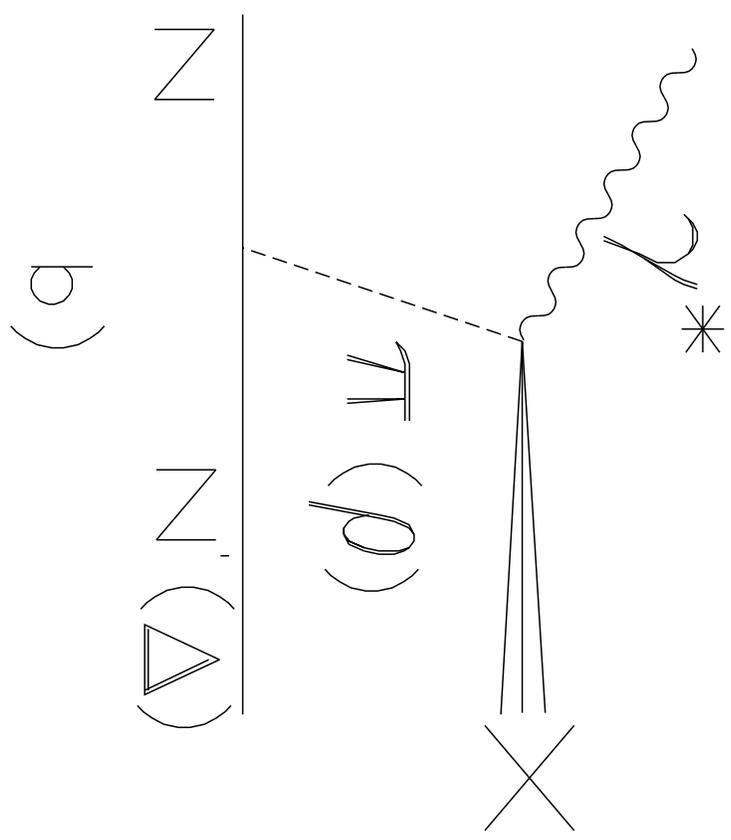
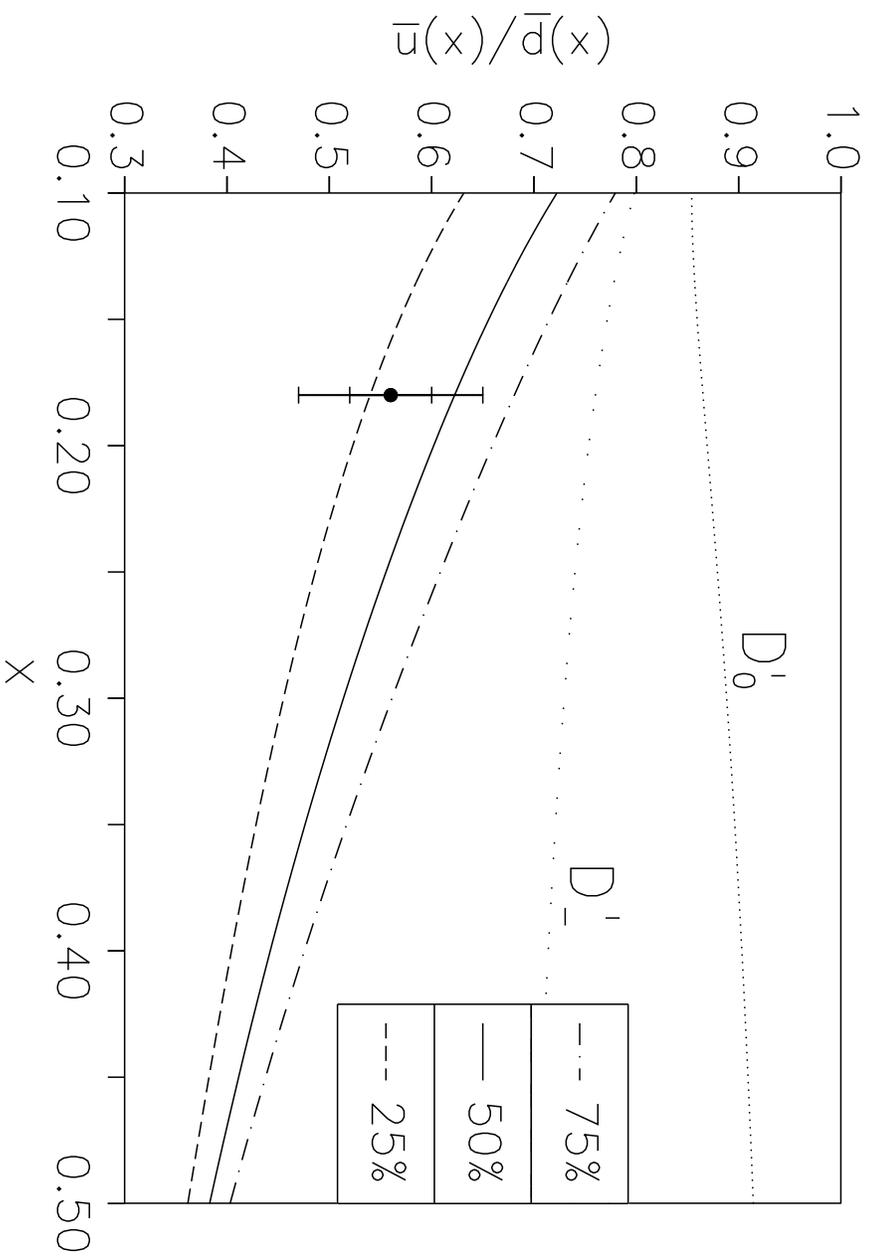
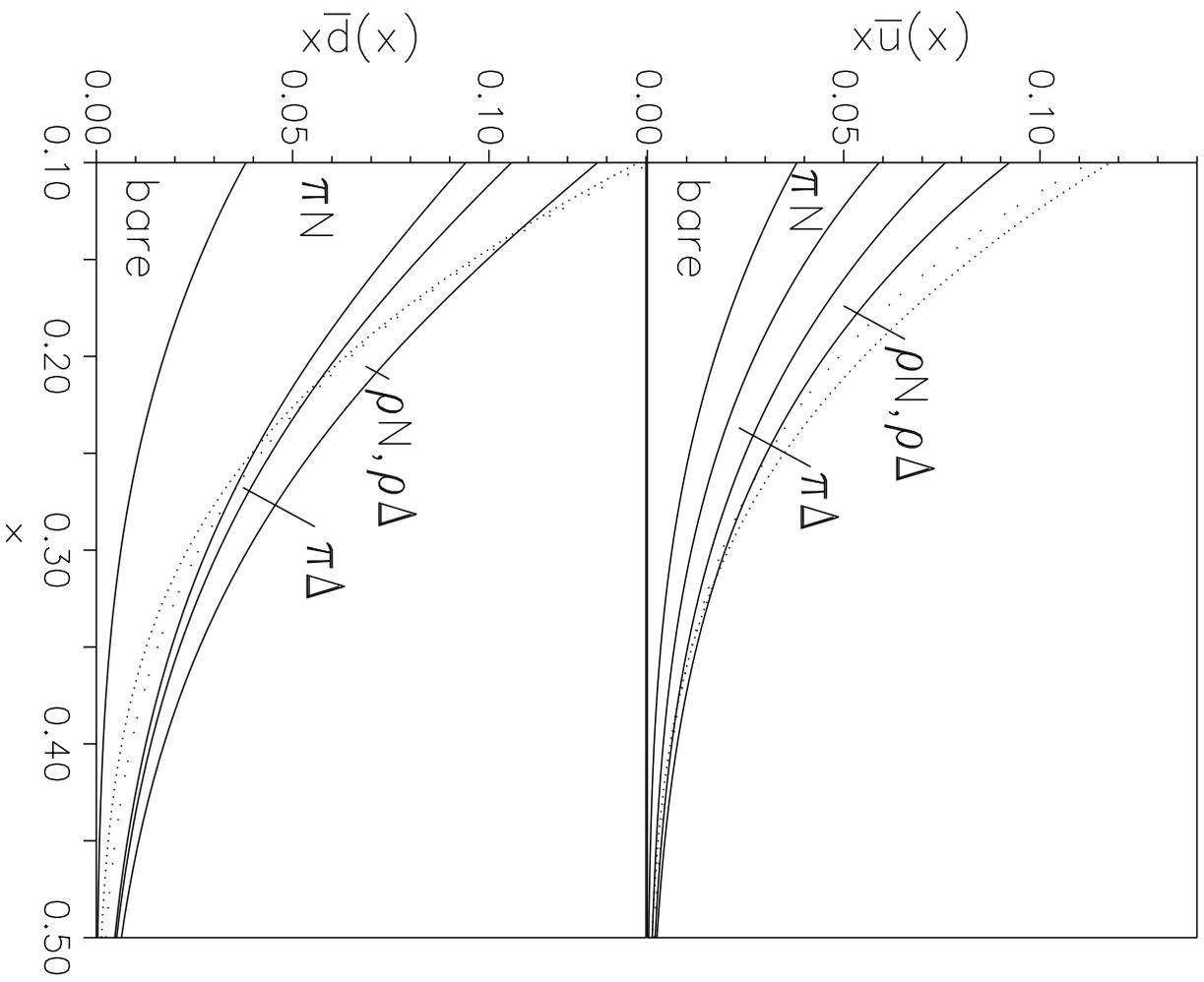
\end{document}